\documentclass[12pt,preprint]{emulateapj}
\def\apj{{ApJ}}
\def\mnras{{ MNRAS}}

\def\be{\begin{equation}}
\def\ee{\end{equation}}
\def\bea{\begin{eqnarray}}
\def\eea{\end{eqnarray}}

\usepackage{graphicx}

\begin{document}

\title{GRB 131231A: Implications of the GeV emission}

\author{Bin Liu\altaffilmark{1,2}, Wei Chen\altaffilmark{1,3},
    Yun-Feng Liang\altaffilmark{1}, Bei Zhou\altaffilmark{1,4}, Hao-Ning He\altaffilmark{1}, Pak-Hin Thomas Tam\altaffilmark{5}, Lang Shao\altaffilmark{2}, Zhi-Ping Jin\altaffilmark{1}, Yi-Zhong Fan\altaffilmark{1}, and Da-Ming Wei\altaffilmark{1}}
\altaffiltext{1}{Key laboratory of Dark Matter and Space Astronomy, Purple Mountain Observatory, Chinese Academy of Sciences, Nanjing 210008, China;}
\altaffiltext{2}{Department of Physics, Hebei Normal University, Shijiazhuang 050024, China;}
\altaffiltext{3}{School of Physics, Huazhong University of Science and Technology, Wuhan 430074, China;}
\altaffiltext{4}{University of Chinese Academy of Sciences, Yuquan Road 19, Beijing, 100049, China;}
\altaffiltext{5}{Institute of Astronomy and Department of Physics, National Tsing Hua University, Hsinchu 30013, Taiwan.}
\email{liangyf@pmo.ac.cn (YFL), beizhou@pmo.ac.cn (BZ)}
\email{phtam@phys.nthu.edu.tw (PHT), yzfan@pmo.ac.cn (YZF), dmwei@pmo.ac.cn (DMW)}

\begin{abstract}
GRB 131231A was detected by the Large Area Telescope onboard Fermi Space Gamma-ray Telescope.
The high energy gamma-ray ($> 100$ MeV) afterglow emission spectrum is $F_\nu \propto \nu^{-0.54\pm0.15}$ in the first $\sim 1300$ s after the trigger and the most energetic photon has an energy $\sim 62$ GeV arriving at $t\sim 520$ s.
With reasonable parameters of the GRB outflow as well as the density of the circum-burst medium, the synchrotron radiation of electrons or protons accelerated at an external forward shock have difficulty accounting for the data.
The synchrotron self-Compton radiation of the forward shock-accelerated electrons, instead, can account for both the spectrum and temporal behavior of the GeV afterglow emission.
We also show that the prospect for detecting GRB 131231A$-$like GRBs with Cherenkov Telescope Array (CTA) is promising.
\end{abstract}


\setlength{\parindent}{.25in}

\section{Introduction}
The Fermi Space Gamma-ray Telescope (FSGT) was launched in 2008 \citep{Fermi2009Inst}.
In the pre-FSGT era, the most-widely discussed mechanism of generating GeV emission of Gamma-Ray Bursts (GRBs) is the inverse Compton (IC) scattering (e.g., M\'{e}sz\'{a}ros \& Rees 1994; Chiang \& Dermer 1999; Dermer et al. 2000; Fan \& Piran 2008; Razzaque 2013). The possibly dominant role of synchrotron radiation of the shock-accelerated electrons in producing GeV emission has not been seriously-investigated/noticed until Nov. 2008, two months before the official release of the first Fermi-LAT GRB data (the first report was on GRB 080916C by Abdo et al. 2009a), by \citet{Zou2009}. The subsequent modeling of the Fermi-LAT GRB GeV emission data with the synchrotron radiation of the forward shock electrons can be found in a series of works \citep[e.g.,][]{Kumar2009, Gao2009, Ghisellini2010, He2011, Ackermann2013, Liang2014} and now the synchrotron radiation has been widely taken to be the leading mechanism to interpret the GRB GeV emission. Nevertheless, the IC should play a non-ignorable role in producing $>10$ GeV afterglow emission \citep{Zou2009}.
The physical reason is that due to the energy loss caused by the synchrotron radiation of the shock-accelerating electrons, these particles can not reach an energy higher than $\gamma_{\rm e,M}m_{\rm e}c^2$, where $\gamma_{\rm e,M}\sim 10^{8}B^{-1/2}$ is the so-called maximal Lorentz factor of the electrons and $B$ is the strength of the magnetic field in the emitting region \citep{Cheng1996}. Correspondingly, the observed synchrotron radiation is not expected to exceed $\epsilon_{\rm syn,M}\sim 0.1~\Gamma/(1+z)~{\rm GeV}$, where $\Gamma$ is the bulk Lorentz factor of the emitting region and $z$ is the redshift of the GRB. In the early afterglow phase, usually we have $\Gamma \lesssim 100$ and hence $\epsilon_{\rm syn,M}\sim 10~{\rm GeV}(\Gamma/100)/(1+z)$. This is a quite general argument, which is valid as long as the emitting electrons are shock-accelerated \citep{Cheng1996,ZM2001,Piran2010}.
Motivated by such a fact, \citet{xue2009} suggested that the $\gamma-$rays more energetic than tens-GeV, if detected in the afterglow of some relatively ``nearby" luminous GRBs at a time $t>$a few hundred seconds, would favor the IC scattering origin.

Among GRBs detected by FSGT \citep[e.g.,][]{Abdo2009a,Abdo2009,Zhang2011,Tam2012,latgrbcatalog}, a remarkable one is GRB 130427A, the most powerful explosion of all GRBs at redshift $z\leq 0.5$ \citep{Fermi2014,Fan2013,Xu2013}. Its emission above 100 MeV lasted about one day and four photons are at energies greater than 40 GeV.
Regardless of the interpretation of the early
multi-wavelength emission data, the simple but robust constraint by $\epsilon_{\rm syn,M}$ strongly disfavors the synchrotron radiation origin of the $>10$ GeV afterglow photons \citep{Fan2013,Wang2013,Liu2013,Fermi2014}. The discovery of a hard spectral GeV-TeV radiation component in the afterglow phase is strongly in support of the IC radiation model (Tam et al. 2013; c.f. Kouveliotou et al. 2013; Perley et al. 2014).

In this work we analyze the Fermi-LAT public data of GRB 131231A \citep{gcn15640} and report the discovery of a hard spectrum component in the first $\sim 10^{3}$s (see section 3) and then we discuss its interpretation and implication, in particular the detection prospect by Cherenkov Telescope Array (CTA) in section 4.

\section{GRB 131231A}
The Fermi Gamma-ray Burst Monitor (GBM) triggered on GRB~131231A~(trigger 410157919) at
04:45:16.08 UT (T$_0$) on Dec 31 2013.
The detection at high enough peak flux in the GBM detectors initiated an Autonomous Repoint Request (ARR) and Fermi slewed to the GRB location.
The location of GRB 131231A is at a distance of 40$^\circ$ from the
LAT boresight at the time of the trigger and LAT kept it in the field
of view (FoV) for 900s \citep{gcn15640}. It entered the LAT FoV again between 3200s and 7000s after the trigger.

In addition to the Fermi/GBM, Konus-Wind also triggered the burst.
In the GBM data, the light curve consists of a single large peak preceded
by a smaller peak which resulted in the trigger. Its duration ($T_\mathrm{90}$) is about 31s
\citep{gcn15644}.
The observation of Konus-Wind shows a broad multi-peaked pulse from 13s before and 35s after it was triggered. The fluence of this burst is $1.55\pm0.05\times10^{-4}$ erg/cm$^2$ according to the observation of Konus-Wind from T0 to T0+7.488s in the 20keV-10MeV energy range \citep{gcn15670}, while the GBM team reported that the fluence is $1.400\pm0.001\times10^{-4}$ erg/cm$^2$ in the time interval T0+0.003s to T0+56s, in the 10keV-1000keV energy range \citep{gcn15672}. The time-averaged spectrum (from trigger to 34.303s after) is also well fitted by Band function \citep{1993ApJ...413..281B} with $E_{\rm peak} = 163\pm6$ keV, $\alpha = -1.28\pm0.04$ and $\beta = -2.47\pm0.05$ \citep{gcn15670}.

In the X-ray band, Swift/XRT started to observe it at 52.1ks after the trigger.
The location of the source observed by the Swift/XRT is RA=10.5904 and Dec=-1.6519 with an uncertainty of 3.5 arcsec(radius) at 90$\%$ confidence level and the position is consistently used in our subsequent analysis of LAT data \citep{gcn15648}.
The optical emission was firstly detected by the 1m telescope located on Mt. Nanshan, Xinjiang, China at T0+7.91 hr
after the Fermi trigger and then confirmed by some other telescopes \citep{gcn15641}. The redshift has been measured to be $\sim 0.643$ \citep{gcn15645,gcn15652}.

\section{Fermi-LAT data analysis of GRB 131231A and the results}
The Fermi/LAT data of GRB 131231A we utilized are available at the Fermi Science Support Center (FSSC)\footnote{http://fermi.gsfc.nasa.gov/ssc/data/access/},
and we used the newest Software, Fermi Science Tools version v9r32p5 package in our analysis.
We selected the events with ``evclass=2'' (source class) from 100MeV to 300GeV within 15$^\circ$ around the circle center ($10.5904^{\circ}$, $-1.6519^{\circ}$) from FSSC. To reduce the contamination of the gamma-ray photons from earth limb, we disregarded all the photons with the zenith angle greater than $100^{\circ}$. Since the time intervals we select are all short, we didn't run \emph{gtmktime} in our analysis procedure, as is usually done in GRB analysis.

\begin{figure}
\centering
\includegraphics[angle=0,scale=0.35,width=0.5\textwidth,height=0.3\textheight]{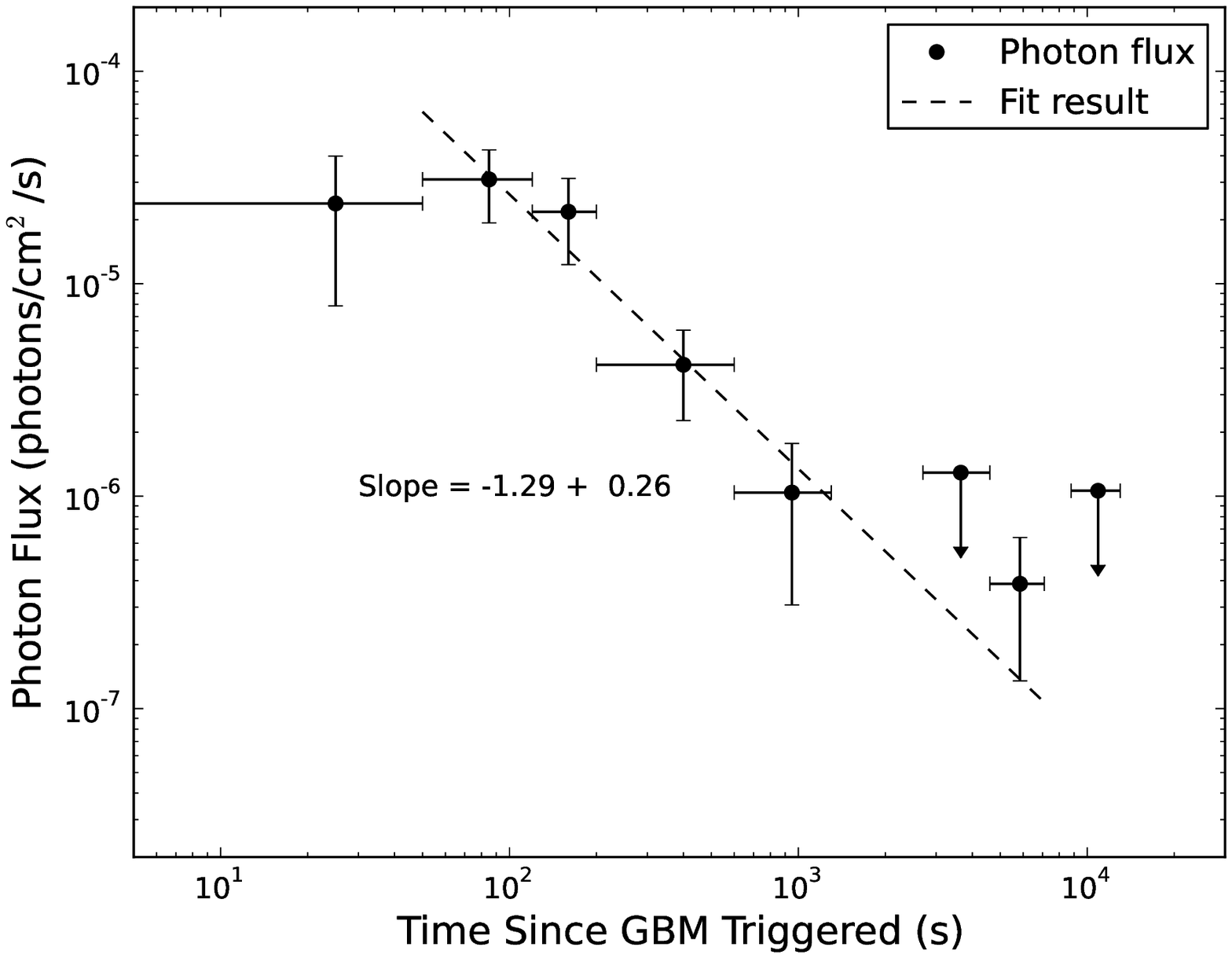}
\includegraphics[angle=0,scale=0.35,width=0.5\textwidth,height=0.3\textheight]{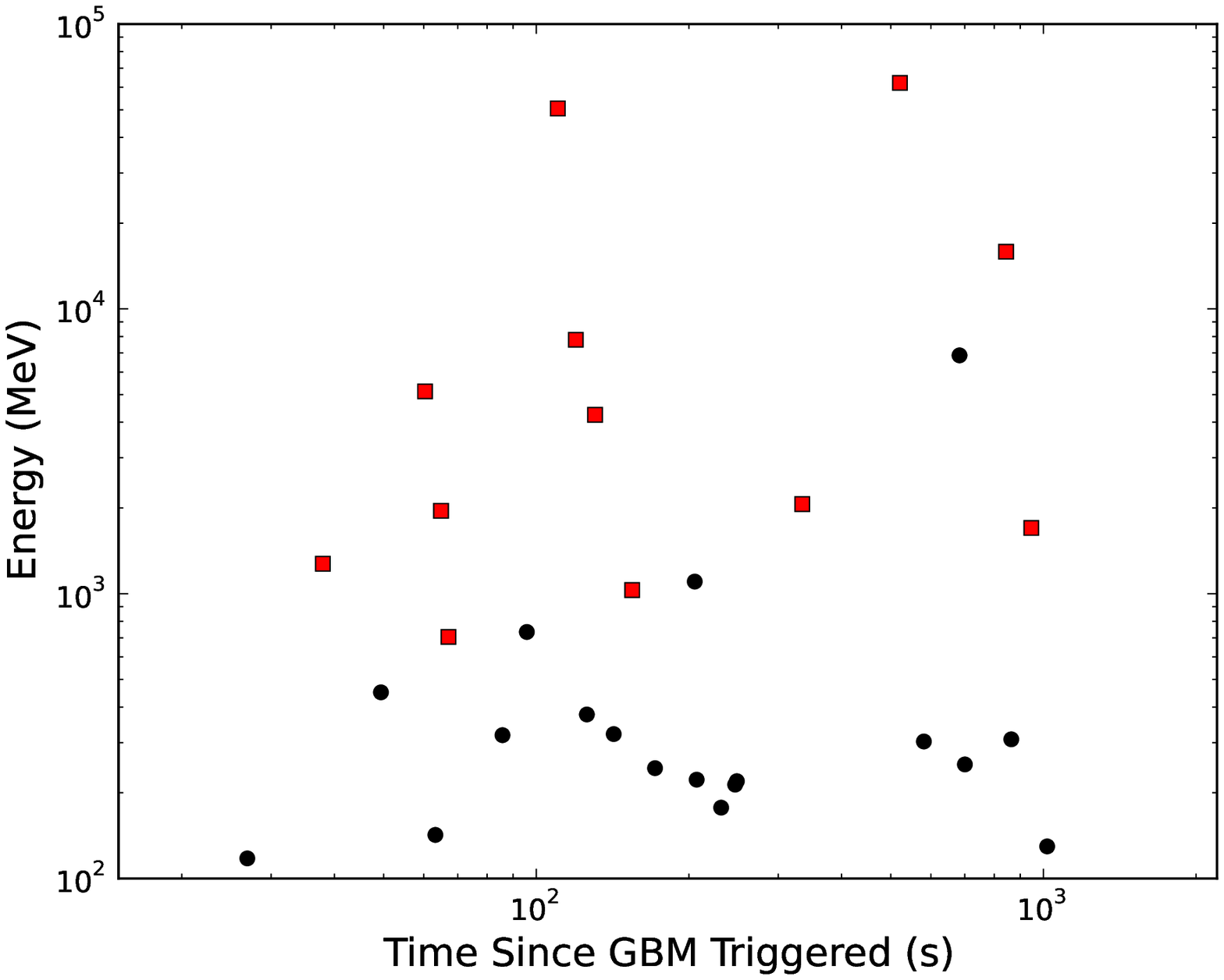}%
\hfill
\caption{\emph{upper panel:} Light curve of the high energy range from 0.1-100GeV emission for GRB131231A.
The best fit for flux variation result (dashed line) shows that the flux decays following a power law with index of $-1.29$. \emph{bottom panel:} Energy of each photon and its arrival time. Here only photons with $P \ge 0.5$ are shown, where $P$ represents the probability of a photon being associated with GRB131231A. The red squares represent those photons with $P \ge 0.997$.}
\label{fig:lt}
\end{figure}

\subsection{The high energy light curve of GRB$~$131231A}

To construct the light curve of GRB131231A, we performed unbinned maximum likelihood fits.
The photons were divided into 8 time intervals to acquire sufficient statistics in each interval.
Since some photons come from the diffuse background or point sources and the time spans in each bin is short,
we added the ``Galactic diffuse" (``gll\_iem\_v05.fits") , ``isotropic diffuse" (``iso\_source\_v05.txt") components and all the point sources within 20$^\circ$ from the GRB position into our background model, keeping the parameters of all the point sources fixed.
We assumed a single power law spectrum for GRB$~$131231A with its ``normalization" and ``spectral index" being allowed to vary.

\begin{figure}
\centering
\includegraphics[angle=0,scale=0.350,width=0.5\textwidth,height=0.35\textheight]{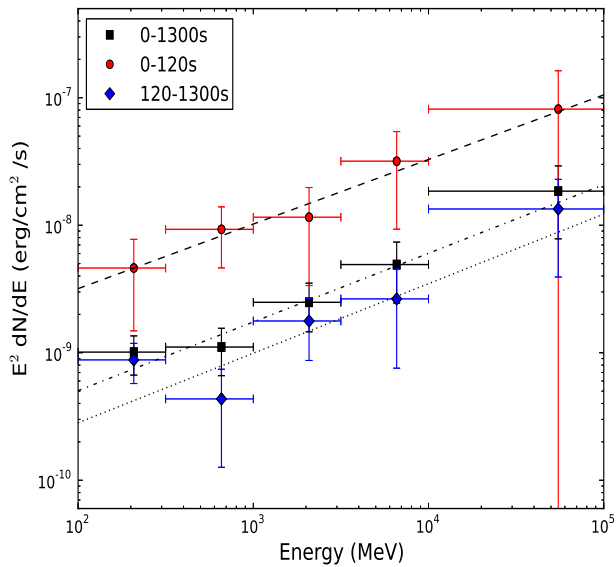}%
\hfill
\caption{Spectrum of GRB131231A in the first 1300s (interval A, denoted by squares), and in two sub epochs: $T_\mathrm{0}$ to $T_\mathrm{0}+120$s (interval B, denoted by circles) and $T_\mathrm{0}+120$s to $T_\mathrm{0}+1300$s (interval C, denoted by diamonds). The dash-dotted, dashed and dotted lines are the best fit spectra. The index of the best fitted line are 0.54$\pm$0.15, 0.51$\pm$0.21, 0.55$\pm$0.28 for interval A, B and C respectively.}
\label{fig:sed}
\end{figure}

The results of our analysis are shown in Fig.\ref{fig:lt}. We further fit the light curve with a single power law excluding the first time interval since it is before the peak flux time.
The best fit temporal index is $-1.29\pm0.26$, which is similar to other Fermi-LAT GRBs \citep{Zhang2011,latgrbcatalog}.

\subsection{Spectral analysis of the prompt and afterglow phase}
As far as photon statistics allow, we attempt to build the spectral energy distributions (SEDs) from the prompt phase to the afterglow phase.
The criteria we used to select the data are similar to those used to construct the light curve except that we split all data into two time intervals (i.e. \rm T$_0$ to T$_0$+120s and T$_0$+120s to T$_0$+1300s) and five energy bins. Attempts to use more time intervals may in principle give a more detailed time-evolution of the gamma-ray spectrum, as suggested by~\citet{Fermi2014} for the case of GRB~130427A. However, due to the limited photon statistics especially above a few GeV, we only chose two time intervals with enough statistics to derive a general feature of the spectrum.
The method and the models we used in this procedure are the same as the last section when constructing the light curve.
We plot the resulting data in Fig.\ref{fig:sed}.
We then fit the five spectral points with a single power law and the index is $0.51\pm0.21$ for the first time interval and $0.55\pm0.28$ for the second time interval. The fit of all data in the first $\sim 1300$ s gives a power-law index $0.54\pm0.15$. These indices are consistent with those obtained by maximum likelihood analysis using all 100~MeV to 100~GeV photons.
The energy spectrum from 100~MeV to 100~GeV is very hard, which is rare among Fermi-LAT GRBs \citep{Zhang2011,latgrbcatalog},  except for GRB 100728A, which has photon index, $\Gamma\sim-1.4$ during an X-ray flare (Abdo et al., 2011, He et al. 2012). Although some bursts in the first Fermi GRB catalog may have spectrum almost as hard as that of GRB 131231A, especially for GRB 110120A and GRB 110428A \citep{latgrbcatalog}, these bursts have a lower detection significance as well as a smaller photon number associated with the respective GRB.

We present the probability of each photon in this data in the bottom panel of Fig.\ref{fig:lt}, adopting the \emph{gtsrcprob} tool.
As shown in the Fig.\ref{fig:lt} there are three photons with energy more than 10 GeV: a 16 GeV photon arrived at $\rm T_0+844$s, a 51 GeV photon at $\rm T_0+110$s and a 62 GeV photon at $\rm T_0+521.3$s.
The probability that each of these $>$10~GeV photons is associated with the GRB is $>$99.7\%.
With a redshift $z\approx 0.643$, the highest gamma-ray has an energy $\approx 102$ GeV in the burst's frame. The implication of such energetic photons will be discussed later.

\subsection{Comparison between GRB 131231A and GRB 130427A}
It would be interesting to compare the high energy emission properties with those of GRB 130427A. In GRB 130427A, four gamma-rays above 40 GeV were collected \citep{Fermi2014,Fan2013} while in GRB 131231A just two gamma-rays are found to be so energetic. We note that: (i) the isotropic-equivalent kinetic energy of the outflow of GRB 131231A is smaller than that of GRB 130427A, as suggested by the less luminous prompt emission of GRB 131231A (whose isotropic 0.1 $-$ 100 GeV energy release is $6.9\pm3.6\times 10^{52}$ erg but $2.1\pm0.3\times10^{53}$ erg for GRB 130427A); (ii) GRB 131231A is at a redshift $z\approx 0.643$, which is a factor of two larger than GRB 130427A ($z\approx0.34$). The extragalactic background light (EBL), which reduces the chance of a $\sim$100~GeV photon to arrive to the Earth by photon-photon pair production, may play a moderate role here\footnote{EBL effects may not be very severe even in the case of GRB~131231A, as there is no clear signature of a spectral cutoff at the highest end in the spectrum, albeit with limited photon statistics}. Taking these two factors into account, GRB~131231A is actually comparable to GRB~130427A in the number of very energetic photons that it emits.

In GRB 130427A a hard spectral component ($\Gamma\sim-1.4$) above a few GeV was found during the afterglow phase \citep{Tam2013}, while in GRB 131231A the hard spectrum emerges at energy as low as $\sim 100$ MeV. Such a fact suggests that either the synchrotron radiation of GRB 131231A is considerably less-efficient than that of GRB 130427A or the IC radiation of GRB 131231A is much more-efficient than that of GRB 130427A.

\section{Discussion}
GRB 131231A was detected by FSGT but not {\it Swift/BAT}, and the early X-ray and optical observations are not available. Here we adopt the GeV emission to constrain the afterglow model. In the electron synchrotron radiation model, a spectrum as hard as $F_\nu \propto \nu^{-0.54\pm 0.15}$ and the flux decline $t^{-1.29\pm 0.26}$ can be roughly accounted if (i) the energy distribution index of injected electrons is as hard as $p\sim 1.2$ and the GeV band is above both the typical synchrotron radiation frequency ($\nu_{\rm m}$) and the cooling frequency ($\nu_{\rm c}$)\footnote{In the case of $1<p<2$, a reliable calculation of the forward shock emission is still not available and two groups of primary approaches have been developed by Bhattacharya (2001) and Dai \& Cheng (2001) and by Xue et al. (2009b), respectively.}, or alternatively (ii)  the energy distribution index of injected electrons is normal with $p\sim 2.2$ and the GeV band is between $\nu_{\rm m}$ and $\nu_{\rm c}$. The synchrotron radiation spectrum of shock-accelerated electrons, however, is expected to have a cutoff at the energy (i.e.,  eq.(8) of Fan et al. (2013))
\begin{eqnarray}
\epsilon_{\rm syn,M} &\approx & \left\{%
\begin{array}{ll}
    11~{\rm GeV}~E_{\rm k,54}^{1/8}n_{-2}^{-1/8}t_{2.7}^{-3/8}({1+z\over 1.64})^{-5/8}, & \hbox{ISM;} \\
     10~{\rm GeV}~E_{\rm k,54}^{1/4}A_{*,-2}^{-1/4}t_{2.7}^{-1/4}({1+z \over 1.64})^{1/4}, & \hbox{wind;} \\
\end{array}%
\label{eq:Syn_limit}
\right.
\end{eqnarray}
where $E_{\rm k}$ is the isotropic-equivalent kinetic energy of the outflow, $n$ is the number density of the medium, and $A_*$ is the wind parameter. The convenience $Q_{\rm x}=Q/10^{\rm x}$ has been adopted.
For reasonable $E_{\rm k}$ and $n$ (or $A_*$), the $\sim 62$ GeV $\gamma-$ray arriving at $t\sim 521$ s (see Fig.\ref{fig:lt}) cannot be accounted for. We then conclude that the synchrotron radiation origin is disfavored unless the electrons are accelerated by mechanisms that operate on timescales faster than the Larmor timescale.

Eq.(\ref{eq:Syn_limit}) does not hold any longer if the high energy emission is from shock-accelerated protons rather than electrons \citep{Razzaque2010}. Intriguingly, the synchrotron radiation of slowly-cooling protons has a spectrum $F_\nu \propto \nu^{-({\bar{p}}-1)/2}$, which is well consistent with the observation, where $\bar{p} \sim 2.2$ is the distribution index of shock-accelerated protons.
The GeV emission flux of the protons can be estimated as
\begin{equation}
F_{\rm GeV,p}
=({m_{\rm e}\over m_{\rm p}})^{(3p-1)/2}({\epsilon_{\rm p}\over \epsilon_{\rm e}})^{p-1}F_{\rm \nu, max}({\nu_{\rm GeV}\over \nu_{\rm m}})^{-(p-1)/2},
\end{equation}
where
$\epsilon_{\rm p}$ ($\epsilon_{\rm e}$) is the fraction of shock energy given to the protons (electrons) and $F_{\rm \nu, max}$ is the maximal specific fluxes of the forward shock electrons. With eqs.(2-3) of \citet{FP06}, it is straightforward to show that $E_{\rm k}\sim 10^{57}$ erg is needed to reproduce the observed GeV flux $\sim 10^{-9}~{\rm erg~s^{-1}~cm^{-2}}$ at $t\sim 10^{3}$ s. Though such energies make possible the acceleration of ultra-high energy cosmic rays in the forward shock region, such a huge $E_{\rm k}$ seems to be unrealistic.

Below we turn to the synchrotron self-Compton (SSC) radiation model \citep[e.g.,][]{Meszaros1994,Chiang1999,Dermer2000,Sari2001,ZM2001,Fan2008}.
As summarized in eqs.(52-53) of \citet{FP08}, the characteristic frequencies governing the SSC spectrum are
$\nu_{\rm m,ssc}\propto t^{-9/4}$ in the interstellar medium (ISM) case and $\propto t^{-2}$ in the free wind medium, and $\nu_{\rm c,ssc}\propto t^{-1/4}$ in the ISM case and $\propto t^{2}$ in the free wind medium. Please note that $\nu_{\rm m,ssc}\approx 2\gamma_{\rm m}^{2}\nu_{\rm m}$ and $\nu_{\rm c,ssc}\approx 2\gamma_{\rm c}^{2}\nu_{\rm c}$, where $\nu_{\rm m}$ is the typical synchrotron radiation frequency of the forward shock electrons and $\gamma_{\rm m}$ is the minimum Lorentz factor of the shock-accelerated electrons, and $\nu_{\rm c}$ is the so-called cooling frequency of the forward shock electrons and $\gamma_{\rm c}$ is the cooling Lorentz factor of the forward shock \citep{Sari2001}. On the other hand, the maximum SSC flux can be estimated as $F_{\rm \nu_{\rm max},ssc} \propto t^{1/4}$ in the ISM case and $\propto t^{-1}$ in the wind case (see eq.(54) of Fan \& Piran (2008) for a summary).

In order to account for the observed spectrum $F_\nu \propto \nu^{-0.54\pm 0.15}$, the electrons are likely either in fast cooling phase (i.e., $\nu_{\rm c,ssc}<\nu<\nu_{\rm m,ssc}$) or in slow cooling phase (i.e., $\nu_{\rm m,ssc}<\nu<\nu_{\rm c,ssc}$) if $p$ is not sizeably larger than 2. In the former we would expect a temporal behavior $F_\nu \propto t^{1/8}$ in the ISM case and $\propto t^{0}$ in the wind case, both are inconsistent with the observed decline $\propto t^{-1.29\pm 0.26}$ (see the Fig.\ref{fig:lt}). For the electrons in the slow cooling phase, we have $F_\nu \propto t^{\rm (11-9p)/8}$ in the ISM case and $\propto t^{-p}$ in the wind case. Clearly the ISM model is in agreement with both the spectrum data and the temporal behavior for $p\approx 2.3$ while the wind medium model is not.

Due to the lack of simultaneous optical and X-ray afterglow emission at $t<5\times 10^{4}$ s, it is impossible to reliably infer the shock parameters. Even so, a rough estimate on the importance of the IC radiation (i.e., the so-called Compton parameter $Y$) is achievable. In the SSC model $\nu_{\rm m,ssc}<100$ MeV at $t\sim 100$ s and $\nu_{\rm c,ssc}\gtrsim 100$ GeV at $t\sim 10^{3}$ s are needed. As a conservative estimate of $Y$, we assume that the synchrotron and SSC radiation components contribute equally at $\nu \sim 100$ MeV. Note that both $\nu_{\rm c}$ and $\nu_{\rm m}$ are likely to be well below $\sim 100$ MeV, and hence the synchrotron radiation flux at $\sim 100$ MeV should drop with time as $\propto t^{(2-3p)/4}\propto t^{-1.25}$, very similar to that of the detected decline of high energy afterglow emission. Therefore we have $Y \gtrsim \nu_{\rm c,ssc}F_{\nu_{\rm c,ssc}}/\nu_{\rm c}F_{\nu_{\rm c}}\sim (\nu_{\rm c,ssc}/100~{\rm GeV})^{0.35}(\nu_{\rm c}/100~{\rm eV})^{0.15}$, implying that the forward shock electrons were likely mainly cooled by the synchrotron radiation photons rather than by the shock-generated magnetic field.

\begin{figure*}
\centering
\includegraphics[angle=0,scale=0.35,width=0.5\textwidth,height=0.35\textheight]{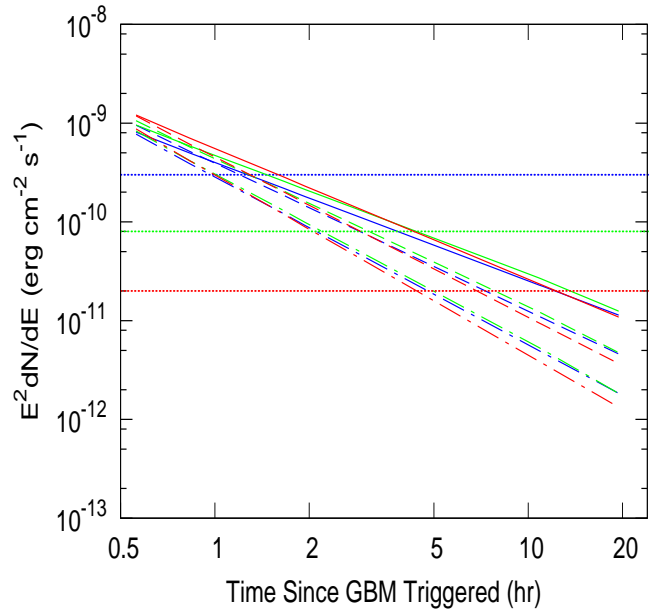}%
\hfill
\caption{The detectability of the tens-GeV emission of GRB 131231A-like bursts for CTA with an exposure of half an hour. The dashed lines are the sensitivities of CTA with an exposure of half an hour, where the red, green and blue lines are for $\nu=(100,~40,~25)$ GeV, respectively. The solid, dashed, dash-dotted lines are the emission $\nu F_\nu$ as a function of time for $p=2.3$, $2.5$ and $2.7$ (i.e., the declines are $\propto t^{-1.3}$, $t^{-1.5}$ and $t^{-1.7}$), respectively.}
\label{CTA}
\end{figure*}

CTA is a project to build a new generation ground-based gamma-ray instrument in the energy range extending from tens-GeV to above 100 TeV. Compared to current generation of Imaging Atmospheric Cherenkov Telescope arrays, CTA has a factor of $5-10$ improvement in sensitivity in the energy range of ${\rm 100 GeV-10 TeV}$ and the extension to energies about 25 GeV \citep{CTAC2011,Kakuwa2012}, CTA will be very suitable to observe the high energy afterglow emission of GRBs. That's why now we estimate the detection prospect of GRB 131231A by CTA-like detectors.
For $t>10^{3}$s, we would expect $\nu_{\rm c,ssc}\approx 100~{\rm GeV}~(t/10^{3}~{\rm s})^{-1/4}$. The SSC radiation spectrum is thus $F_\nu \propto \nu^{-(p-1)/2}$ for $\nu_{\rm m,ssc}<\nu<\nu_{\rm c,ssc}$ and $F_\nu \propto \nu^{-p/2}$ for $\nu>\nu_{\rm c,ssc}$.
For a given $p$, we can estimate the flux, say, at $\nu=(25,~40,~100)$ GeV, respectively (see lines shown in Fig.\ref{CTA}). As a conservative estimate, we assume that the exposure time is half an hour, for which the CTA sensitivities as a function of photon energy can be found in Fig.24 of \citet{CTAC2011}. One can find in Fig.\ref{CTA} that the $\sim 25$ GeV emission might be a bit challenging to be detected by CTA unless the observation started very early, for example, within one hour after the GRB trigger. The $\sim 40$ GeV emission could be detectable in a few hours after the trigger while the detection prospect of $\sim 100$ GeV emission is even more promising.

In view of these facts, we conclude that (1) for the hard spectrum of the high energy ($100~{\rm MeV}-100~{\rm GeV}$) afterglow
emission of GRB 131231A, the model of synchrotron radiation of the shock-accelerated electrons or protons is less likely while the synchrotron self-Compton scattering origin is favored; (2) the fireball was in slow-cooling phase and the circum-burst medium had a constant density profile; (3) CTA could significantly detect tens of GeV afterglow emission of GRB 131231A-like events and then play an important role in revealing the underlying physical process(es).

\section*{Acknowledgments}
We thank the referee for insightful comments. This work was supported in part by 973 Programme of China under grants 2013CB837000 and 2014CB845800, National Natural Science of China under grants 11273063 and 11361140349, China Postdoctoral science foundation under grant 2013T60569, and the Foundation for Distinguished Young Scholars of Jiangsu Province, China (No. BK2012047).  YZF is also supported by the 100 Talents programme of Chinese Academy of Sciences. PHT is supported by the National Science Council of the Republic of China (Taiwan) through grant NSC101-2112-M-007-022-MY3.

\begin{deluxetable}{cccccc}
\tablewidth{0pt}
\tabletypesize{\footnotesize}
\tablecaption{Result of unbinned maximum likelihood analysis in each time interval}
\tablehead{
\colhead{Time(s)} & \colhead{Flux\tablenotemark{a}} & \colhead{Fluence\tablenotemark{b}} & \colhead{Photon Index\tablenotemark{c}} & \colhead{TS value} & \colhead{Npred\tablenotemark{d}}
}				
\startdata
0--50&(2.39$\pm$1.60)$\times$10$^{-5}$&(1.25$\pm$8.32)$\times$10$^{-6}$&-2.02$\pm$0.54&16.03&3.00\\
50--120&(3.10$\pm$1.17)$\times$10$^{-5}$&(1.92$\pm$1.50)$\times$10$^{-5}$&-1.33$\pm$0.20&94.97&8.00\\
120--200&(2.18$\pm$9.51)$\times$10$^{-6}$&(3.49$\pm$7.18)$\times$10$^{-6}$&-1.79$\pm$0.31&36.96&5.92\\
200--600&(4.15$\pm$1.89)$\times$10$^{-6}$&(6.42$\pm$8.56)$\times$10$^{-6}$&-1.58$\pm$0.26&28.45&7.11\\
600--1300&(10.39$\pm$7.32)$\times$10$^{-7}$&(1.03$\pm$1.28)$\times$10$^{-5}$&-1.18$\pm$0.37&33.26&3.92\\
2700--4600&1.29$\times$10$^{-6}$&(1.56$\pm$1.61)$\times$10$^{-7}$&0&2.13&1.48\\
4600--7100&(3.87$\pm$2.52)$\times$10$^{-7}$&(0.90$\pm$5.12)$\times$10$^{-6}$&-2.07$\pm$0.53&11.6&3.93\\
8800--13000&1.06$\times$10$^{-6}$&(0.15$\pm$1.46)$\times$10$^{-5}$&0&3.99&1.07\\
\enddata
\tablenotetext{a}{In the unit of $photons{\cdot}cm^{-2}s^{-1}$; values without uncertainty are upper limits.}
\tablenotetext{b}{In the unit of $ergs{\cdot}cm^{-2}$.}
\tablenotetext{c}{Index values are listed only for those intervals with TS$>$9.}
\tablenotetext{d}{Predicted photon number given by the unbinned likelihood analysis}
\label{tab:lt}
\end{deluxetable}

\end{document}